\documentclass{nature}
\usepackage{graphicx}
\usepackage{float}
\usepackage{placeins}
\usepackage{gensymb}
\usepackage{bm}
\usepackage{amssymb}
\usepackage{amsmath}
\usepackage{times}
\usepackage{color}
\usepackage{float}

\usepackage{braket}
\usepackage{epstopdf}
\usepackage{mathtools}
\usepackage{mathrsfs}
\usepackage{caption}
\usepackage{subfig}

\allowdisplaybreaks

\usepackage{newfloat}

\DeclareFloatingEnvironment[name={Supplementary Figure},fileext=lof]{suppfigure}

\title{Collective atomic scattering and motional effects in a dense coherent medium}

\author{S. L. Bromley$^{1}$,  B. Zhu$^{1}$,  M. Bishof$^{1,}\footnote{Current Address: Physics Division, Argonne National Laboratory, Argonne, Illinois 60439, USA}$  , X. Zhang$^{1,}\footnote{Current Address: International Center for Quantum Materials, School of Physics, Peking University, Beijing 100871, China}$, T. Bothwell$^{1}$,  J. Schachenmayer$^{1}$,  T. L. Nicholson$^{1,}\footnote{Current Address: Center for Ultracold Atoms, Massachusetts Institute of Technology, Cambridge, MA 02139, USA}$, R. Kaiser$^{2}$, S. F. Yelin$^{3,4}$, M. D. Lukin$^{4}$ , A.M. Rey$^{1\S}$, \& J. Ye$^{1\S}$}

\begin{document}

\maketitle

\begin{affiliations}
\item JILA, NIST, and Department of Physics, University of Colorado, 440 UCB, Boulder, CO 80309, USA
\item Universit$\acute{e}$ de Nice Sophia Antipolis, CNRS, Institut Non-Lin$\acute{e}$aire de Nice, UMR 7335, F-06560 Valbonne, France
\item Department of Physics, University of Connecticut, Storrs, CT 06269, USA
\item Department of Physics, Harvard University, Cambridge, MA 02138, USA

${}^\S$ Corresponding authors. E-mail:arey@jilau1.colorado.edu and Ye@jila.colorado.edu.
\end{affiliations}

\begin{abstract}

We investigate collective emission from coherently driven ultracold $\bm{^{88}}$Sr atoms. We perform two sets of experiments, using a strong and weak transition that are insensitive and sensitive, respectively, to atomic motion at one microKelvin. We observe highly directional forward emission with a peak intensity that is enhanced, for the strong transition, by $\bm{>}$$\bm{10^{3}}$ compared to that in the transverse direction. This is accompanied by substantial broadening of spectral lines. For the weak transition, the forward enhancement is substantially reduced due to motion. Meanwhile, a density-dependent frequency shift of the weak transition ($\bm{\sim}$ 10\% of the natural linewidth) is observed. In contrast, this shift is suppressed to $\bm{<}$1\% of the natural linewidth for the strong transition. Along the transverse direction, we observe strong polarization dependences of the fluorescence intensity and line broadening for both transitions. The measurements are reproduced with a theoretical model treating the atoms as coherent, interacting, radiating dipoles.

\end{abstract}

\section{Introduction}
Understanding interactions between light and matter in a dense atomic medium is a long-standing problem in physical science\cite{Gross1982, Andreev1980} since the seminal work of Dicke\cite{Dicke1954}.   In addition to their fundamental importance in optical physics, such interactions play a central role in enabling a range of new quantum technologies including optical lattice atomic clocks\cite{Bloom2014} and quantum networks\cite{Kimble08}.

The key ingredient in a dense sample is dipole-dipole interactions that arise from the exchange of virtual photons with dispersive and radiative contributions, and their relative magnitude varies between the near-field and far-field regimes.  The dispersive (real) part is responsible for collective level shifts and the radiative (imaginary) part gives rise to line broadening and collective superradiant emission\cite{supexp1973,supexp1985,yelinpra}. Intense theoretical efforts have been undertaken over many years to treat the complex interplay between the dispersive and radiative dynamics\cite{Friedberg1974,FriedbergHartman,agarwalana,Lewenstein,HSteudel1978,rzazewski2atom,RuostekoskiPRA1997,scullyshift,lehmberg,dfvjames}. However, experimental demonstrations that provide a complete picture to clarify these interactions have been elusive.

Collective level shifts and line broadening arising from the real and imaginary parts of dipole-dipole interactions have recently been observed in both atomic\cite{CAdams2012,CLS2014,antoine2015,antoine20152,antoine2016} and condensed matter\cite{CLS2009} systems. The modification of radiative decay dynamics at low excitation levels has also been observed using short probe pulses\cite{Havey2013, SrFlash2011,SrFlash2014,Kaiser2015}, and interaction effects were manifested in coherent backscattering\cite{Kaiser1999,Havey2003}. While simple models of incoherent radiation transport have often been used to describe light propagation through opaque media\cite{rmp1998,rmp1999} and radiation trapping in laser cooling of dense atomic samples\cite{Wieman1990}, coherent effects arising from atom-atom interactions, which are necessary to capture correlated many-body quantum behavior induced by dipolar exchange, are beginning to play a central role.  For example, the dipole-dipole interaction is responsible for the observed dipolar blockade and collective excitations in Rydberg atoms\cite{Pfau07, Lukin01,Saffman2010,Pfau2009,Pfau2012,Peyronel2012, Antoine14, Gunter13}; it may also place a limit to the accuracy of an optical lattice clock and will require non-trivial lattice geometries to overcome the resulting frequency shift\cite{Chang2004}.  Previous theoretical efforts have already shown that physical conditions such as finite sample size, sample geometry, and the simultaneous presence of dispersive and radiative parts can play crucial roles in atomic emission\cite{kus,Lewenstein,Friedberg1974,FriedbergHartman,zakowiczsphere,HSteudel1978,Sutherland}.

In this work we use millions of Sr atoms in optically thick ensembles, taking advantage of the unique level structure of Sr to address motional effects, to study these radiative and dispersive parts simultaneously.  We demonstrate that a single, self-consistent, microscopic theory model can provide a unifying picture for the majority of our observations.  These understandings can help underpin emerging applications based on many-body quantum science,  such as lattice-based optical atomic clocks\cite{Bloom2014,Nicholson2015,Katori2015}, quantum nonlinear optics\cite{Peyronel2012}, quantum simulations\cite{igor}, and atomic ensemble-based quantum memories\cite{Rempe2011}.


\section{Results}
\subsection{Experimental Setup}
\begin{figure*}

  \centering
  \includegraphics[width=6cm]{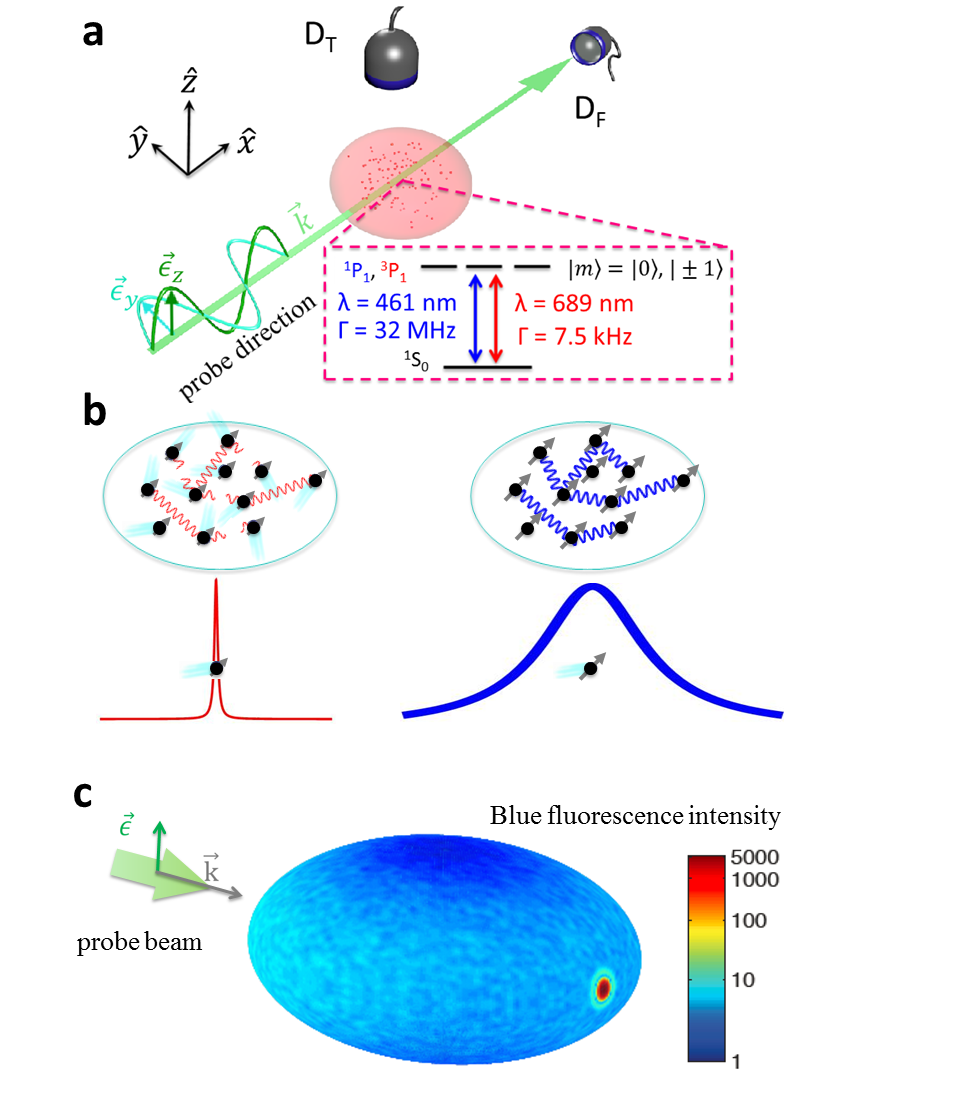}
  \caption[The short caption]{\textbf{(a)} The experimental scheme and concept.  We weakly excite the strontium atoms with a linearly polarized probe beam and measure the fluorescence with two detectors, one in the forward direction, $\hat{x}$, and the other almost in the perpendicular direction, $\hat{z}$.  We probe two different J = 0 to J' = 1 transitions.  The first transition is a $^{1}S_{0}-$$^{1}P_{1}$  blue transition with a natural linewidth of $\mathit{\Gamma} = 32$ MHz and the second is a $^{1}S_{0}-$$^{3}P_{1}$ red transition with $\mathit{\Gamma} = 7.5$ kHz.  \textbf{(b)} In the coherent dipole model photons are shared between atoms. When the Doppler broadened linewidth becomes comparable to the natural linewidth, dephasing must be considered.  At our $\sim 1\mu$K temperatures the Doppler broadening is $\approx 40$ kHz meaning motional effects are important only for the red transition.  \textbf{(c)}  The 3D intensity distribution predicted for a blue probe beam.  The coupled-dipole model predicts a strong $10^{3}$ enhancement of the forward intensity compared to other directions and a finite fluorescence along a direction parallel to the incident polarization. The speckled pattern is due to randomly positioned atoms and can be removed by averaging over multiple atom configurations.  }
\label{fig:setup}
\end{figure*}

Bosonic alkaline-earth atoms with zero nuclear spin have simple atomic structure compared to the more complex hyperfine structure present in typical alkali-metal atoms that complicates the modeling and interpretation of the experimental observations. For example, $^{88}$Sr atoms have both a strong $^{1}S_{0}-$$^{1}P_{1}$ blue transition ($\mathit{\lambda}=461$ nm) and a spin-forbidden weak $^{1}S_{0}-$$^{3}P_{1}$ red transition ($\mathit{\lambda}=689$ nm), with a strict 4-level geometry (Fig.~\ref{fig:setup}(a)).  When the atoms are cooled to a temperature of $\sim$1~$\mu$K, Doppler broadening at 461 nm is about 55 kHz, which is almost three orders of magnitude smaller than the blue transition natural linewidth, $\mathit{\Gamma}$ = 32~MHz. To an excellent approximation atomic motion is negligible for atomic coherence prepared by the 461 nm light. To the contrary, the red transition with a natural linewidth $\mathit{\Gamma}$ = 7.5~kHz is strongly affected by atomic motion. By comparing the behaviors of the same atomic ensemble observed at these two different wavelengths (Fig.~\ref{fig:setup}(b)) we can thus collect clear signatures of motional effects on coherent scattering and dipolar coupling\cite{juhashift,thermal}.

We use the experimental scheme shown in Fig.~\ref{fig:setup}(a) to perform a comprehensive set of measurements of fluorescence intensity emitted by a dense sample of $^{88}$Sr atoms.  The sample is released from the trap and then illuminated with a weak probe laser. We vary the atomic density, cloud geometry, observation direction, and polarization state of the laser field, and we report the system characteristics using three key parameters: the peak scattered intensity, the linewidth broadening, and the line center shift. For example, along the forward and transverse directions we observe different values of intensity and linewidth broadening, as well as their dependence on light polarization (see Fig.~\ref{fig:setup}(c)). We also observe motional effects on the red transition in contrast to the same measurements on the blue transition.

In the experiment, up to 20 million $^{88}$Sr atoms are cooled to $\sim$1~$\mu$K in a two-stage magneto-optical trap (MOT), the first based on the blue transition and the second on the red transition. The atomic cloud is then released from the MOT and allowed to expand for a variable time of flight (TOF), which allows us to control its optical depth and density.  They are subsequently illuminated for 50(100)~$\mu$s with a large size probe beam resonant with the blue (red) transition (Fig.~\ref{fig:forward}(a)). The resulting scattered light is measured with two detectors far away from the cloud (see Fig.~\ref{fig:setup} (a)).  One detector is along the forward direction $\hat{x}$ (detector $D_{\mathrm F}$), and the other transverse direction $\hat{z}$ (detector $D_{\mathrm T}$, offset by $\sim$$10^\circ$). For a short TOF, the atomic cloud is anisotropic and has an approximately Gaussian distribution with an aspect ratio of $R_{x}:R_{y}:R_{z}=2:2:1$ , where $R_{\{ x,y,z \}}$ are r.m.s radii.  We define $OD$ as the on resonance optical depth of the cloud, $OD=\frac{3 N}{2\left( kR_{\perp}\right )^{2}}$, where $R_{\perp}$ depends on the direction of observation with $R_{\perp,\mathrm {T}}=R_{x}=R_{y}$ and $R_{\perp,{\mathrm{F}}}=\left( R_{z}R_{y}\right )^{1/2}$ for the transverse and forward directions respectively, $N$ is the atom number, and ${k}$ is the laser wavevector for the atomic transition (see Supplementary Note 1).

\subsection{Forward Observations}
\begin{figure*}
  \centering
	\includegraphics[width=18cm]{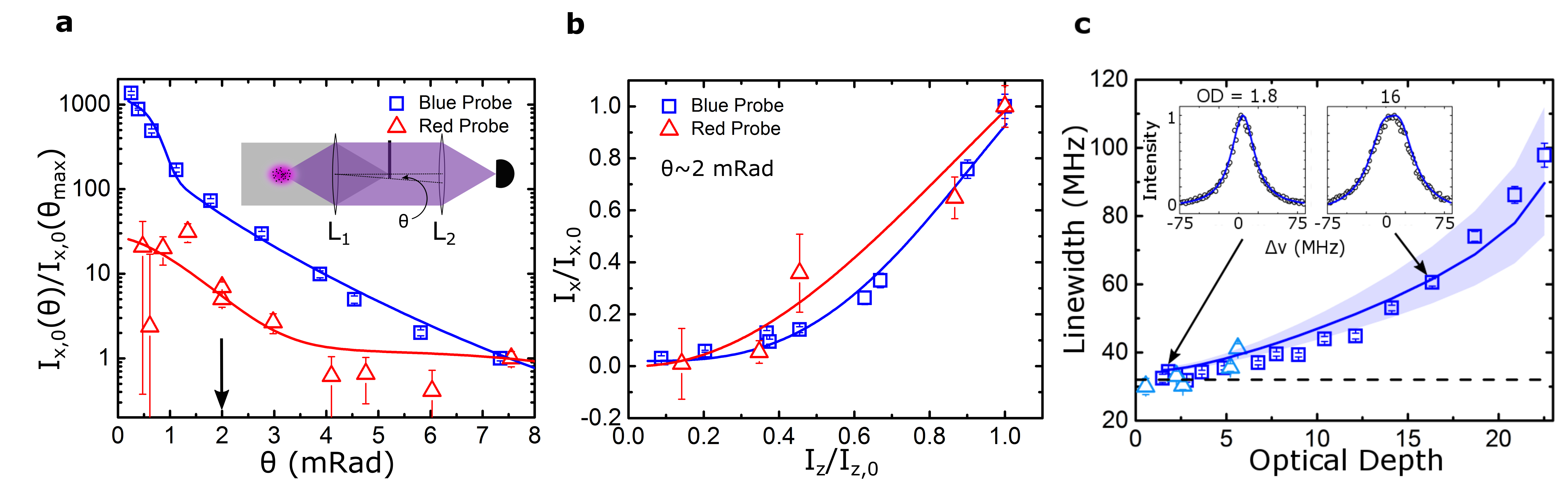}
	\caption{\textbf{Forward Scattering. (a)} Comparison of forward scattering intensity versus angle using a red and blue probe beam.  We use the setup shown in the inset to block the probe beam.  After interacting with the atoms the probe beam is focused using a lens, which also collimates the fluorescence from the atoms.  We block the probe beam using a beam stopper, which we translate perpendicular to the probe beam to change the angular range of fluorescence collected by the detector, characterized by the angle ($\theta$) between $\hat{x}$ and the edge of the beam stopper (see Methods). The measured intensity, $I_{x,0}\left (\theta \right )$, for each probe beam is normalized to the intensity at $\theta_{\mathrm{max}}=7.5$ mRad.  The dephasing caused by motion reduces the forward intensity peak for the red transition.  \textbf{(b)}  Comparison of intensity in the forward direction, $I_{x}$, versus intensity in the transverse direction, $I_{z}$. Both are varied by changing $N$.  All measurements are made at $\theta=2$ mRad (arrow in (a)) and normalized to the intensity, $I_{x,0}$, for the atom number used in (\textbf{a}).  \textbf{(c)} Linewidth broadening in the forward direction measured by scanning the blue probe beam detuning, $\mathit{\Delta}$, across resonance.  Example lineshapes for different optical depths ($OD$s) are shown in the inset.  Two different atom numbers are used, $N=1.7(2) \times 10^{7}$ (blue squares) and $N/4$ (cyan triangles).  The dashed line represents $\mathit{\Gamma}$ for reference.  All solid curves are based on the full theory of coupled dipoles and the band in (c) is for a $\pm 20\%$ atom number uncertainty. All errorbars are for statistical uncertainties.  }
	\label{fig:forward}
\end{figure*}

The coherent effect manifests itself most clearly in the forward direction (Fig.~\ref{fig:forward}). To separate the forward fluorescence from the probe beam, we focus the probe with a lens (L$_{1}$) after it has passed through the atomic cloud and then block it with a beam stopping blade, which can be translated perpendicular to the probe beam (Fig.~\ref{fig:forward}(a) inset). The same lens (L$_1$) also collimates the atomic fluorescence so that it can be imaged onto $D_{\rm F}$. The position of the beam stopper can be used to vary the angular range of collected fluorescence, characterized by the angle ($\theta$) between $\hat{x}$ and the edge of the beam stopper (see Methods). Using the maximum atom number available in the experiment, the measured intensity $I_{x,0}\left (\theta \right )$ is normalized to that collected at $\theta_{\mathrm{max}}=7.5$ mRad. Both the blue (square) and red (triangle) transition results are displayed in Fig.~\ref{fig:forward}(a). For the blue transition we observe a thousand-fold enhancement of the normalized intensity for $\theta<$ 0.5 mRad. We note that this enhancement as a function of $\theta$ far exceeds the diffraction effect by the beam stopper. The latter effect becomes significant only when $\theta$ $\lesssim 0.1$~mRad, and is also suppressed in our differential measurement scheme which compares measurements with and without atoms.  The enhancement is also present for the red transition, but it is reduced by nearly two-orders of magnitude at small $\theta$ due to the motional effect.  On the other hand, the wider angular area of enhancement is attributed to the longer wavelength of the red transition.  The forward intensity strongly depends on the atom number. In Fig.~\ref{fig:forward}(b) we present measurements of the forward intensity $I_{x}$ versus the transverse intensity $I_{z}$ at a fixed $\theta=2$ mRad for different atom numbers. The intensities are normalized to those obtained at the peak atom number as used in Fig.~\ref{fig:forward}(a). To the first order approximation, the transverse fluorescence intensity scales linearly with the atom number. Hence, the forward intensity of both the blue and red transitions scales approximately with the atom number squared.

In the forward direction we have also investigated the linewidth broadening of the blue transition as a function of the atomic $OD$. By scanning the probe frequency across resonance we extract the fluorescence linewidth, which is found to be determined primarily by the $OD$ of the cloud (open squares in Fig.~\ref{fig:forward}(c)). For the range of $0<OD<20$ the lineshape is Lorentzian (see insets); however, the observed lineshape starts to flatten at the center for $OD$$>$20. We have also varied the atom number by a factor of four, and to an excellent approximation the linewidth data is observed to collapse to the same curve when plotted as a function of $OD$ (open triangles).

\subsection{Transverse Observations}
\begin{figure*}
  \centering
  	\includegraphics[width=18cm]{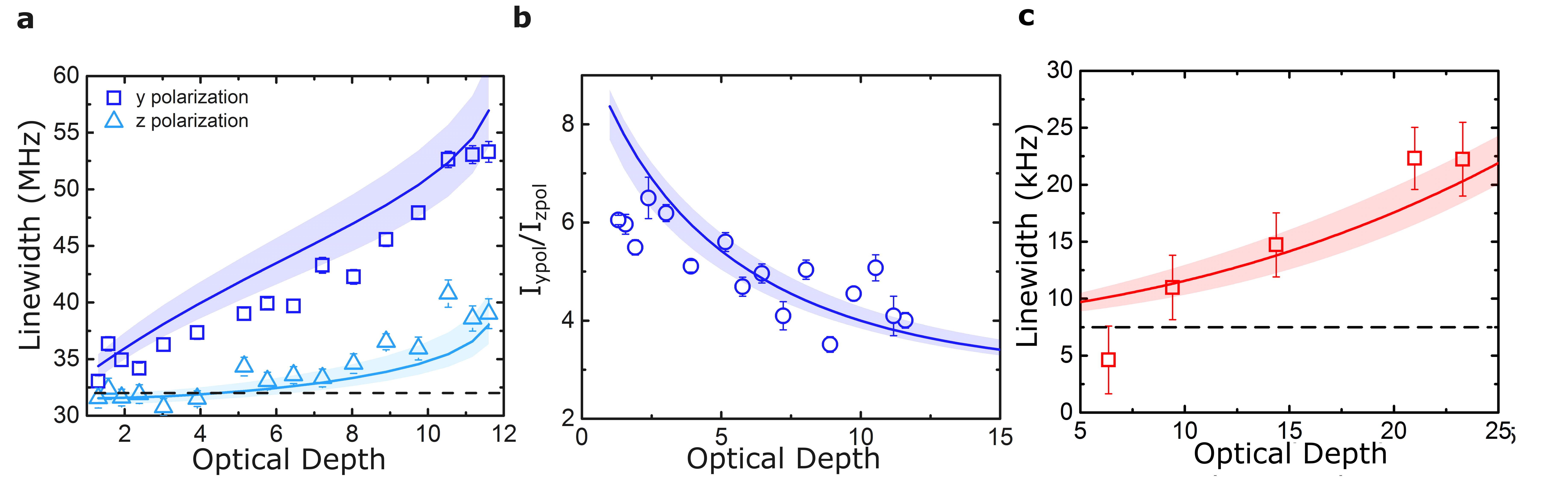}
	\caption{\textbf{Transverse Scattering.  (a)}  Linewidth broadening for the blue transition in the transverse direction for $\hat{y}$-polarization (open squares) and $\hat{z}$-polarization (open triangles).  \textbf{(b)} Intensity ratio, $I_{\mathrm{ypol}}/I_{\mathrm{zpol}}$, of $\hat{y}$-polarization to $\hat{z}$-polarization measured in the transverse direction when a blue probe beam is used.  For low optical depths single particle scattering is dominant, and for single particle scattering almost zero intensity is predicted for $\hat{z}$-polarized fluorescence as this polarization points directly into the detector.  \textbf{(c)}  Linewidth broadening for the red transition in the transverse direction for $\hat{y}$-polarized light showing a similar trend to the blue transition.  This transition is more sensitive to magnetic fields so a large magnetic field is applied to probe only the m = 0 to m' = 0 transition.  All solid curves are based on the full theory of coupled dipoles and the band in (c) is for a $\pm 20\%$ atom number uncertainty. All errorbars are for statistical uncertainties.  }
	\label{fig:transverse}
\end{figure*}

For independent emitters the forward fluorescence should have no dependence on the probe beam polarization; however, the transverse fluorescence (along $\hat{z}$) should be highly sensitive to the probe polarization, and it is even classically forbidden if the probe is $\hat{z}$-polarized. However, multiple scattering processes with dipolar interactions can completely modify this picture by redistributing the atomic population in the three excited magnetic states and thus scrambling the polarization of the emitted fluorescence. Consequently, even for a $\hat{z}$-polarized probe there should be a finite emission along $\hat{z}$ (see Fig.~\ref{fig:setup}(c)), with an intensity that increases with increasing $OD$. Our experimental investigation of the fluorescence properties along the transverse direction is summarized in Fig.~\ref{fig:transverse}. Under the same $OD$ along $\hat{z}$, the $\hat{y}$-polarized probe beam (square) gives rise to a much more broadened lineshape for the blue transition than the $\hat{z}$-polarized probe beam does (triangle), as shown in Fig.~\ref{fig:transverse}(a). Meanwhile, the peak intensity ratio of $ I_{\rm ypol}/I_{\rm zpol}$ decreases significantly with an increasing $OD$, indicating the rapidly rising fluorescence with a $\hat{z}$-polarized probe when $OD$ increases (Fig.~\ref{fig:transverse}(b)). For the red transition, the existence of Doppler broadening requires the lineshape data to be fitted to a Voigt profile. With the Doppler linewidth $\mathit{\Delta}_{\rm D}$ fixed from the thermal velocity measured in free expansion, the Voigt profile determines the line center as well as the Lorentzian linewidth with the Gaussian linewidth determined by the temperature. Fig.~\ref{fig:transverse}(c) displays the Lorentzian linewidth obtained with a $\hat{y}$-polarized red probe showing a strong increase of the linewidth with $OD$.

\subsection{Spectral Broadening and Shift}
\begin{figure*}
  \centering
  	\includegraphics[width=0.5\textwidth]{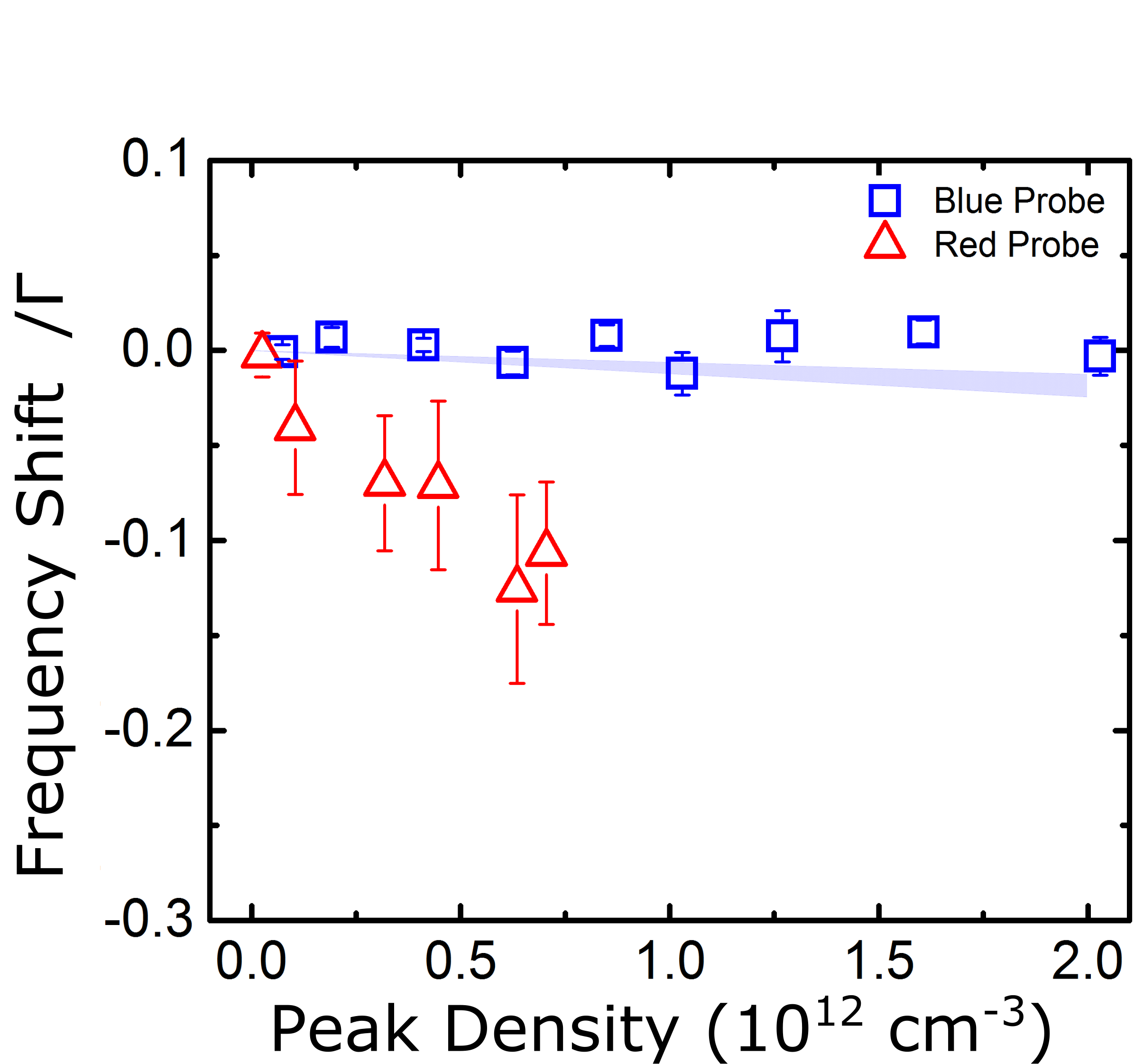}
	\caption{\textbf{Frequency Shift.} Comparison of frequency shift normalized to the corresponding natural linewidth for the blue and red transitions. The blue frequency shift is consistent with zero to 0.004 of $\mathit{\Gamma}$ at an atomic density of $10^{12}$ cm$^{-3}$. The red shift, on the other hand, shows more than 0.1$\mathit{\Gamma}$ at densities up to $0.7 \times 10^{12}$ cm$^{-3}$. }
	\label{fig:fshift}
\end{figure*}

To a good approximation, the dependence of the linewidth on $OD$ along the forward and transverse directions (for the classically allowed $\hat{y}$ polarization in the single scattering limit) is similar. However, due to the anisotropic aspect ratio of the cloud, for the same TOF, the $OD$ is lower along $\hat{z}$ than along $\hat{x}$. This is responsible for the smaller broadenings measured along $\hat{z}$ than along $\hat{x}$. The classically forbidden polarization direction, on the other hand, exhibits a different scaling with $OD$, which is understandable given that the emission in this case comes only from multiple scattering events with dipolar interactions.  The transverse linewidth broadening for the red transition is similar to that of the blue, and it does not depend sensitively on motional effects.  This behavior is in stark contrast to another important observation: the shift of the transition center frequency. Fig.~\ref{fig:fshift} contrasts the linecenter frequency shift observed for $^{1}S_{0}-$$^{1}P_{1}$ (square) and $^{1}S_{0}-$$^{3}P_{1}$ (triangle, with original data reported in Ref.~[\citenum{Ido2005}], see Supplementary Fig. 1). The blue transition frequency shift is consistent with zero at the level of 0.004$\mathit{\Gamma}$ using an atomic density of $10^{12}$ cm$^{-3}$. However, the measured density shift for the red transition (normalized to the transition linewidth) is more than one order of magnitude larger. This density-related frequency shift significantly exceeds the predicted value based on general S-matrix calculations of $s$-wave collisions\cite{Ido2005} (2.18$\times$10$^{-10}$ Hz cm$^{3}$ if the unitary limit is used).

\subsection{Theory Model}
Before we turn to a microscopic model to obtain a full and consistent understanding of all these related experimental observations, we note that semiclassical models\cite{resonantmultiple} treating the atomic cloud as a continuous medium of an appropriate refractive index can give an intuitive explanation of the linewidth broadening in the forward direction.  Classically, an incoming electric field is attenuated as it propagates through the medium according to the Beer--Lambert law, and the forward fluorescence intensity is determined by the same mechanism. This simple semiclassical model recovers the linear dependence of the forward width for small $OD$, and predicts a nonlinear dependence of the linewidth for large $OD$ and a flattening of the line center.  However, we find that this semiclassical approach cannot provide explanations for most aspects of the experimental observations.

The full microscopic model builds on a set of coherently coupled dipoles. Here, each 4-level atom is treated as a discrete radiating dipole located at a frozen position but coupled with retarded dipole radiation, and it is driven with a weak incident laser beam. The atomic ensemble follows the Gaussian distribution observed in the experiment with the appropriate aspect ratio. By solving the master equation in the steady state we find that the coherence, $b_j^\alpha={\rm Tr}[|g\rangle\langle e_j^\alpha|\hat{\rho}]$, of atom $j$, located at ${\bf r}_j$ is modified by other atoms as\cite{dfvjames,refindex,vector,dalibard,bienaime2011,bienaime2013,opticaltheo,cord2002}:
 \begin{equation}
b_j^\alpha=\frac{\Omega^\xi e^{i{\bf k}\cdot{\bf r}_j}}{2 (\mathit{\Delta}^\alpha+i \frac{\mathit{\Gamma}}{2})}\delta_{\alpha,\xi}+\mathit{\Gamma}\sum_{\alpha',m\neq j}\frac{G_{\alpha,\alpha'}({\bf r}_j-{\bf r}_m)}{(i\mathit{\Delta}^\alpha-\frac{\mathit{\Gamma}}{2})}b_m^{\alpha'}.\label{eq:eom}
\end{equation} Here, $|g\rangle={}^1S_0$, $|e^\alpha\rangle$ corresponds to the three excited states of $^1P_1$ or $^3P_1$, with $\alpha\in \{ x,y,z \}$ representing the Cartesian states. Also, $\hat{\rho}$ is the reduced density matrix of the atoms, and $\delta_{\gamma,\gamma'}$ is the Kronecker Delta. The driving laser's linear polarization state $\xi$ is along $\hat{y}$ or $\hat{z}$, with wavevector ${\bf k}$ along $\hat{x}$, Rabi frequency $\Omega^\xi$ and detuned by $\mathit{\Delta}^\alpha$ from the $|g\rangle\rightarrow|e^\alpha\rangle$ transition.  The function $G_{\alpha,\alpha'}({\bf r})$ accounts for the retarded pairwise dipolar interactions and is given by\cite{igor,dfvjames,dalibard} $ G_{\alpha,\alpha'}({\bf r})=-i\frac{3}{4}\frac{e^{ikr}}{kr}[(\delta_{\alpha,\alpha'}-\hat{r}_\alpha\hat{r}_{\alpha'})+(\delta_{\alpha,\alpha'}-3\hat{r}_\alpha\hat{r}_{\alpha'})(\frac{i}{kr}-\frac{1}{(kr)^2})]$. The fluorescence intensity $I({\bf r}_{\rm s})=\langle\hat{\bf E}^{(+)}({\bf r}_{\rm s})\hat{\bf E}^{(-)}({\bf r}_{\rm s})\rangle$, detected at position ${\bf r}_{\rm s}$, can be determined\cite{dfvjames,lehmberg} as a function of $b_j^\alpha$,
\begin{eqnarray}
I({\bf r}_{\rm s})&\approx&\frac{\mathit{\Gamma}^2}{4\mu^2r_{\rm s}^2}\sum_{j,m}[{\bf b}_j\cdot{\bf b}_m^*-({\bf b}_j\cdot\hat{\bf r}_{\rm s})({\bf b}_m^*\cdot\hat{\bf r}_{\rm s})]e^{i{\bf k}_{\rm s}\cdot({\bf r}_j-{\bf r}_m)},\label{eq:int}
\end{eqnarray} with $\mu$  the atomic transition dipole moment, and  ${\bf k}_{\rm s}=k\hat{\bf r}_{\rm s}$.

\section{Discussions}
To understand the forward enhancement we first consider non-interacting atoms under the zeroth order approximation.  The atomic coherence is driven only by the probe field that imprints its phase and polarization onto the atoms: $b_j^{\alpha(0)}=\frac{i\delta_{\alpha,\xi}\Omega^\xi e^{i{\bf k}\cdot{\bf r}_j}/2}{i\mathit{\Delta}-\mathit{\Gamma}/2}$. The corresponding intensity, $I({\bf r}_{\rm s})=\frac{\mathit{\Gamma}^2}{4\mu^2r_{\rm s}^2}\frac{|\Omega^\xi|^2}{4(\mathit{\Delta}^2+\mathit{\Gamma}^2/4)}(N+N^2e^{-|{\bf k}_{\rm s}-{\bf k}_0|^2R_{\perp,{\rm F}}^2})$
has a Lorentzian profile.  It also exhibits an $N^2$ scaling and an enhanced forward emission lobe, with an angular width given by the ratio between the transition wavelength and the transverse size of the sample $\mathit{\Delta}\theta\sim 1/(k R_{\perp,{\rm F}})$. The forward lobe reflects the constructive interference of the coherently emitted radiation stimulated by the laser. Outside the coherent lobe the constructive interference is quickly reduced due to the random position of atoms\cite{opticaltheo,Eberly,SrFlash2014}.  The longer wavelength of the red transition corresponds to a wider angular width of the forward lobe for the red fluorescence.

Simple considerations can also give rise to a qualitative understanding of atomic motion-related effects on forward enhancement. Again for the red transition, the Doppler effect introduces random phases accumulated by $\delta\phi\sim kv/\mathit{\Gamma}$. Here, $v$ is the thermal velocity. The dephasing reduces coherent photon emission and gives rise to a net suppression of the forward emission intensity. The suppression becomes stronger with $\mathit{\Delta}_{\rm D}/\mathit{\Gamma}$, with $\mathit{\Delta}_{\rm D}=\sqrt{\frac{ k_{\rm B} T}{8  m\mathit{\lambda}^2 \ln{2}}}$ the Doppler width. Such a suppression is clearly observed for the red transition.

To address the linewidth broadening we now consider atoms coupled by dipolar interactions, which tend to emit collectively in an optically dense cloud. The collective emission manifests itself with a broader fluorescence linewidth. Moving to the first order approximation, we note that the atomic coherence acquires contributions not only from the probe beam but also from the surrounding atoms, with $b_j^{\alpha}\sim b_j^{\alpha(0)}+b_j^{\alpha(1)}$. Here,
$b_j^{\alpha(1)}=\frac{i\Omega^\xi\mathit{\Gamma}/2}{(i\mathit{\Delta}-\mathit{\Gamma}/2)^2}K_{\alpha,\xi}^je^{i k x_j}$, and $K_{\alpha,\alpha'}^j=\sum_{m\neq j}G_{\alpha,\alpha'}({\bf r}_j-{\bf r}_m)e^{i k(x_m-x_j)}$. For a relatively dilute cloud with average inter-particle distance $\bar{r}\gg 1/k$, the far-field interactions dominate, so higher order terms beyond $1/r$ can be neglected. Dipolar interactions modify the fluorescence lineshape, with consequences of both a frequency shift that depends on the cloud peak density $n_0$, and a line broadening that is proportional to $OD$: $\mathit{\Delta}\to \mathit{\Delta}+\overline{\mathit{\Delta}}$ and $\mathit{\Gamma}\to \mathit{\Gamma}+\overline{\mathit{\Gamma}}$, with $\overline{\mathit{\Delta}}=-\frac{3\sqrt{2}\pi n_0k^{-3}}{16}\mathit{\Gamma}$ and $\overline{mathit{\Gamma}}=\frac{ OD}{4}\mathit{\Gamma}$. Thus, the first order approximation provides an intuitive picture about the role of dipolar effects on the lineshape.

However, in a cloud with an increasingly large $OD$, dipolar interactions are stronger and multiple scattering processes become relevant. The first order perturbative analysis then breaks down\cite{Pellegrino2014, Yelin99,Yelin2012}. The full solution of equation~(\ref{eq:eom}) based on the coherent coupled-dipole model becomes necessary to account for multiple scattering processes (see Methods). The first signatures arise from the forward fluorescence intensity, where its naive $N^2$ scaling is reduced with an increasing atom number as a consequence of multiple scattering processes. This effect is observed in both red and blue calculations, and it is expected to be more pronounced on the red transition due to its longer wavelength. However, atomic motion leads to a lower effective $OD$, which tends to suppress multiple scattering processes and thus helps to partially recover the collective enhancement. The solid lines in Figs.~\ref{fig:forward}(a) and (b) represent such quantitative theory calculations for both transitions, which agree with experiment.

Meanwhile, for the linewidth broadening observed in the forward direction, it becomes evident that the scaling of the linewidth vs. $OD$ turns nonlinear at large values of $OD$. The experimental data falls within the shaded area in Fig.~\ref{fig:forward}(c), which represents the full solution with a 20\% uncertainty in the experimental atom number.  Multiple scattering processes are also key to the explanation of the measured fluorescence along the transverse direction, especially for the classically forbidden polarization $\hat{z}$. Indeed, for both intensity and linewidth broadening observed in the transverse direction, under either $\hat{y}$ or $\hat{z}$ probe polarization, the full model (shown as shaded areas in both Fig.~\ref{fig:transverse}(a) and (b)) reproduces well the experimental results on $^{1}S_{0}-$$^{1}P_{1}$.  Taking into account motional dephasing (see Supplementary Note 2), the transverse broadening for $^{1}S_{0}-$$^{3}P_{1}$ is also well reproduced as shown in Fig.~\ref{fig:transverse}(c).

So far, we have shown the observed effects on linewidth and fluorescence intensity are uniquely determined by $OD$. However, following the arguments discussed above, the frequency shift arising from the dipolar coupling is expected to scale with atomic density, $\left|\overline{\mathit{\Delta}}\right|/ \mathit{\Gamma}\propto n_0k^{-3}$, which includes both the collective Lamb shift and the Lorentz-Lorenz shift\cite{juhashift, freqshift}.  For our experimental density this effect is $\lesssim~10^{-3}$, which is consistent with the observed frequency shift for the blue transition (Fig.~\ref{fig:fshift}). (Note that the role of multiple scattering processes is to further suppress this frequency shift mechanism\cite{juhashift}). In contrast, for the red transition, the measured density shift (normalized to $\mathit{\Gamma}$) is significantly larger than what is predicted from the current treatment of interacting dipoles; it is also much bigger than the unitarity limit of $s$-wave scattering. Qualitatively, we expect that as the atoms move and approach each other, the long lived ground-excited state coherence in the red transition can be significantly modified by the collisional process and open higher partial wave channels.  We can thus expect a larger collisional phase shift. This process can be further complicated by atomic recoil, light forces, and Doppler dephasing\cite{Julienne1996}.

We have shown that a coherent dipole model describes light scattering in a dense atomic medium with collective effects and multiple scatterings. The model captures the quantitative features of the experimental observations.  Motional effects, as manifested in dephasing, can be captured in the model as well. Our results provide useful guides for further developments of optical atomic clocks and other applications involving dense atomic ensembles.

\section{Methods}
\subsection{Coherent dipole model}
Here we present the derivation of Eq. (1). The fundamental   assumption  is to treat the  atoms as frozen during the interrogation. This is an excellent approximation if $\hbar \mathit{\Gamma}$ is much faster than other energy scales in the problem. The latter condition  is always satisfied in the case of the blue transition. For the  $J=0$ to $J=1$ configuration exhibited by ${}^{88}$Sr, we can label the $J=0$ ground state as $|g\rangle$ and the excited $J=1$ states  using a  Cartesian basis  $|e^z\rangle=|0\rangle$ , $|e^x\rangle=(|-1\rangle -|+1\rangle)/\sqrt{2 }$, $|e^y\rangle=i (|-1\rangle+ |+1\rangle)/\sqrt{2}$ . Here the $|0,\pm1\rangle$ states are the standard angular momentum ones. In the Cartesian basis the  vector transition operator for the $j$ atom located at ${\bf r}_j$ can be written as $\hat{\bf b}_j^{\alpha}= \hat{x} \hat{b}_j^x + \hat{y} \hat{b}_j^y +\hat{z} \hat{b}_j^z $ . Here  $\hat{b}_j^\alpha=|g\rangle_j\langle e^\alpha|$. In  this basis   the master equation  governing the evolution of the reduced density matrix  of the  $N$    atom ensemble, $\hat{\rho}$, in the presence of an an incident laser beam with linear polarization $\xi$, can be written as \cite{dfvjames}:
\begin{eqnarray}
\frac{d\hat{\rho}}{dt}&=&-\frac{i}{2}\sum_{j,\alpha}\mathit{\Delta}^{\alpha}[\Omega_j\hat{b}_j^{\alpha\dagger}+\Omega_j^{*}\hat{b}_j^\alpha,\hat{\rho}]-i\sum_{\mathclap{\substack{j,m\neq j\\\alpha,\beta}}}[g_{jm}^{\alpha\beta}\hat{b}_j^{\alpha\dagger}\hat{b}_j^\beta,\hat{\rho}]\nonumber\\&&\!+i\!\sum_{j,\alpha}\mathit{\Delta}^{\alpha}[\hat{b}_j^{\alpha\dagger}\hat{b}_j^\alpha,\hat{\rho}]\!+\!\!\!\sum_{\mathclap{\substack{j,m\\\alpha,\beta}}}\!f_{jm}^{\alpha\beta}(2\hat{b}_j^\alpha \hat{b}_m^{\beta\dagger}\!-\!\{\hat{b}_j^{\alpha\dagger}\hat{b}_m^\beta,\hat{\rho}\}),
\end{eqnarray} where $\Omega_j=\Omega^\xi e^{i{\bf k}\cdot{\bf r}_j}$ is the Rabi frequecy of the  incident field,  polarized along $\xi$ ($\hat{\xi}\cdot\vec{k}=0$) and  detuned by  $\mathit{\Delta}^\alpha$ from the atomic transition $|g\rangle\rightarrow|e^\alpha\rangle$. The parameters $g_{jm}^{\alpha\beta}=g_{\alpha,\beta}({\bf r}_j-{\bf r}_m)$ and $f_{jm}^{\alpha\beta}=f_{\alpha,\beta}({\bf r}_j-{\bf r}_m)$ are the components of the elastic and inelastic dipolar interactions between a pair of atoms at position ${\bf r}_j$ and ${\bf r}_m$ respectively, and are given by
\begin{eqnarray}
g_{\alpha,\beta}({\bf r})\!\!&=&\!\frac{3\mathit{\Gamma}}{4}[(y_0(k_0r)\!-\!\frac{y_1(k_0r)}{k_0r})\delta_{\alpha,\beta}\!+\!y_2(k_0r)\hat{r}_\alpha\hat{r}_\beta],\\
f_{\alpha,\beta}({\bf r})\!\!&=&\!\frac{3\mathit{\Gamma}}{4}[(j_0(k_0r)\!-\!\frac{j_1(k_0r)}{k_0r})\delta_{\alpha,\beta}\!+\!j_2(k_0r)\hat{r}_\alpha\hat{r}_\beta],
\end{eqnarray}
where $r=|{\bf r}|=|{\bf r}_j-{\bf r}_m|$, $y_n(x)$, $j_n(x)$ are the spherical Bessel functions of the second and first kind respectively. Here also  $\alpha,\beta=x,y$ or $z$ represent  Cartesian components. The symbol $\delta_{\gamma,\gamma'}$ is the Kronecker Delta. In the low intensity limit, we can project the density matrix into a state space including the ground state $|G\rangle\equiv|g_1,g_2,...g_N\rangle$ and states with only one excitation~\cite{bienaime2011,bienaime2013,opticaltheo} such as $|j\alpha\rangle\equiv|g_1,...e_j^\alpha,...g_N\rangle$. In this reduced space, the relevant equations of motion simplify to
\begin{eqnarray}
\frac{d\rho_{j\alpha,j\alpha}}{dt}&=&-\frac{i}{2}(\Omega_j\delta_{\alpha,\xi}\rho_{G,j\alpha}-\Omega_j^{*}\delta_{\alpha,\xi}\rho_{j\alpha,G})\nonumber\\
&&-i\sum_{m\neq j,\beta}g_{jm}^{\alpha\beta}(\rho_{m\beta,j\alpha}-\rho_{j\alpha,m\beta})\nonumber\\
&&-\sum_{j\neq m,\beta}f_{jm}^{\alpha\beta}(\rho_{m\beta,j\alpha}+\rho_{j\alpha,m\beta})-\mathit{\Gamma}\rho_{j\alpha,j\alpha},\\
\frac{d\rho_{j\alpha,G}}{dt}&=&-\frac{i}{2}(\Omega_j\delta_{\alpha,\xi}\rho_{G,G}-\sum_{m}\Omega_m\rho_{j\alpha,m\xi})\nonumber\\&&-i\sum_{m\neq j,\beta}(g_{jm}^{\alpha\beta}-if_{jm}^{\alpha\beta})\rho_{m\beta,G}\nonumber\\&&+(i\mathit{\Delta}^\alpha-\frac{\mathit{\Gamma}}{2})\rho_{j\alpha,G},\\
\frac{d\rho_{j\alpha,m\beta}}{dt}&=&-\frac{i}{2}(\Omega_j\delta_{\alpha,\xi}\rho_{G,m\beta}-\Omega_m^{*}\delta_{\xi,\beta}\rho_{j\alpha,G})\nonumber\\&&-i(\sum_{l\neq j,\nu}\rho_{l\nu,m\beta}g_{jl}^{\alpha\nu}-\rho_{j\alpha,l\nu}g_{lm}^{\beta\nu})-\mathit{\Gamma}\rho_{j\alpha,m\beta}\nonumber\\&&-(\sum_{l\neq j,\nu}\rho_{l\nu,m\beta}f_{jl}^{\alpha\nu}+\sum_{l\neq m,\nu}\rho_{j\alpha,l\nu}f_{lm}^{\beta\nu}),\\
\frac{d\rho_{G,G}}{dt}&=&-\frac{i}{2}(\sum_{j,\alpha}\Omega_j^{*}\rho_{j\alpha,G}-\Omega_j\rho_{G,j\alpha})+\mathit{\Gamma}(1-\rho_{G,G})\nonumber\\&&+\sum_{\mathclap{\substack{m,j\neq m\\\alpha,\beta}}}f_{jm}^{\alpha\beta}(\rho_{j\alpha,m\beta}+\rho_{m\beta,j\alpha}).
\end{eqnarray} where $ \rho_{G,G}={\rm Tr}[\hat{\rho} |G\rangle\langle G|]$,    $\rho_{j\alpha,m\beta} ={\rm Tr}[\hat{\rho} (\hat{b}_m^{\beta\dagger} \hat{b}_j^\alpha)]$ and $\rho_{j\alpha,G}={\rm Tr}[\hat{b}_j^\alpha \hat{\rho}]$.

Since we are interested in the situation of a weak probe limit, $\Omega^\xi \ll \mathit{\Gamma}$, we expand the density matrix in successive orders of $\Omega^\xi/\mathit{\Gamma}$, $\hat{\rho}=\hat{\rho}^{(0)}+\hat{\rho}^{(1)}+\hat{\rho}^{(2)}+...$, and keep the first order terms. At this  level of approximation $\rho_{G,G}=1$,  $\rho_{j\alpha,m\beta}=0$
 and only the optical coherences  $b_j^\alpha\equiv\rho_{j\alpha,G} $ evolve in time accordingly to  the following set of  linear equations:
\begin{eqnarray}
\frac{db_j^\alpha}{dt}\!&=&\!(i\mathit{\Delta}^\alpha\!-\!\frac{\mathit{\Gamma}}{2})b_j^\alpha\!-\!\frac{i}{2}\Omega_j\delta_{\alpha,\xi}
\!-\!i\!\!\!\sum_{m\neq j,\beta}\!(g_{jm}^{\alpha\beta}\!-\!if_{jm}^{\alpha\beta})b_{m}^\beta.
\end{eqnarray} Here  $G_{\alpha,\beta}({\bf r})=(f_{\alpha,\beta}({\bf r})+ig_{\alpha,\beta}({\bf r}))/\mathit{\Gamma}$. The steady state solution can be obtained by setting $\frac{db_j^\alpha}{dt}=0$ and then solving the subsequent $3N$ linear equations.

\subsection{Measure the  enhancement of forward fluorescence}
To measure the scattered light in the forward direction, we use the setup shown in the inset of Fig. 2(a) to tightly focus and block the probe beam while still collecting most of the atomic fluorescence on the CCD camera. We focus the probe beam, after it interacts with the atoms, to a small spot with $15~\mu$m r.m.s. radius and block it using a beam stopping blade.  We then translate the beam stopper perpendicular to the probe beam by a distance $\mathit{\Delta} x$ from our reference point of $x=0$ which we define as the position of the beam stopper where we see the greatest fluorescence without saturating the CCD camera with the probe beam . As only the forward direction is particularly sensitive to positional changes we convert this change in position to a change in angle simply using $\theta={\rm arctan}\frac{\mathit{\Delta} x}{15 {\rm cm}}$ , where the first lens with a $15$ cm focal length collimates the fluorescence. In numerical calculations, the CCD camera is simulated as a ring area centered around the forward direction and the average intensity collected over the ring is determined. The external radius is set to be large enough to reach the angular region outside the interference cone and the inner angular radius $\theta_{\rm sim}$, simulating the blocking of the signal by the beam stopper, is varied accordingly to the experiment. To account for the difference between $\sigma_{\rm sim}$ and the experiment cloud size, $\theta_{\rm sim}$ is rescaled so that  we satisfy the experimental observation that at $\theta_{\rm max}$ the enhancement factor  drops  to $1$.

\bibliographystyle{naturemag}

\begin{thebibliography}{10}
\expandafter\ifx\csname url\endcsname\relax
  \def\url#1{\texttt{#1}}\fi
\expandafter\ifx\csname urlprefix\endcsname\relax\def\urlprefix{URL }\fi
\providecommand{\bibinfo}[2]{#2}
\providecommand{\eprint}[2][]{\url{#2}}

\bibitem{Gross1982}
\bibinfo{author}{Gross, M.} \& \bibinfo{author}{Haroche, S.}
\newblock \bibinfo{title}{Superradiance: An essay on the theory of collective
  spontaneous emission}.
\newblock \emph{\bibinfo{journal}{Phys. Rep.}} \textbf{\bibinfo{volume}{93}},
  \bibinfo{pages}{301--396} (\bibinfo{year}{1982}).

\bibitem{Andreev1980}
\bibinfo{author}{Andreev, A.~V.}, \bibinfo{author}{Emel'yanov, V.~I.} \&
  \bibinfo{author}{Il'inskii, Y.~A.}
\newblock \bibinfo{title}{Collective spontaneous emission (Dicke
  superradiance)}.
\newblock \emph{\bibinfo{journal}{Sov. Phys. Usp.}}
  \textbf{\bibinfo{volume}{23}}, \bibinfo{pages}{493--514}
  (\bibinfo{year}{1980}).

\bibitem{Dicke1954}
\bibinfo{author}{Dicke, R.~H.}
\newblock \bibinfo{title}{Coherence in spontaneous radiation processes}.
\newblock \emph{\bibinfo{journal}{Phys. Rev.}} \textbf{\bibinfo{volume}{93}},
  \bibinfo{pages}{99--110} (\bibinfo{year}{1954}).

\bibitem{Bloom2014}
\bibinfo{author}{Bloom, B. J., Nicholson, T. L.,  Williams, J. R., Campbell, S. L., Bishof, M., Zhang, X., Zhang, W., Bromley, S. L., \& Ye, J.}
\newblock \bibinfo{title}{A New Generation of Atomic Clocks: Accuracy and Stability at the $10^{-18}$ Level}.
\newblock \emph{\bibinfo{journal}{Nature}} \textbf{\bibinfo{volume}{506}},
  \bibinfo{pages}{71--75} (\bibinfo{year}{2014}).


\bibitem{Kimble08}
\bibinfo{author}{Kimble, H.~J.}
\newblock \bibinfo{title}{The quantum internet}.
\newblock \emph{\bibinfo{journal}{Nature}} \textbf{\bibinfo{volume}{453}},
  \bibinfo{pages}{1023--1030} (\bibinfo{year}{2008}).

\bibitem{supexp1973}
\bibinfo{author}{Skribanowitz, N.}, \bibinfo{author}{Herman, I.~P.},
  \bibinfo{author}{MacGillivray, J.~C.} \& \bibinfo{author}{Feld, M.~S.}
\newblock \bibinfo{title}{Observation of Dicke superradiance in optically
  pumped hf gas}.
\newblock \emph{\bibinfo{journal}{Phys. Rev. Lett.}}
  \textbf{\bibinfo{volume}{30}}, \bibinfo{pages}{309--312}
  (\bibinfo{year}{1973}).

\bibitem{supexp1985}
\bibinfo{author}{Pavolini, D.}, \bibinfo{author}{Crubellier, A.},
  \bibinfo{author}{Pillet, P.}, \bibinfo{author}{Cabaret, L.} \&
  \bibinfo{author}{Liberman, S.}
\newblock \bibinfo{title}{Experimental evidence for subradiance}.
\newblock \emph{\bibinfo{journal}{Phys. Rev. Lett.}}
  \textbf{\bibinfo{volume}{54}}, \bibinfo{pages}{1917--1920}
  (\bibinfo{year}{1985}).

\bibitem{yelinpra}
\bibinfo{author}{Wang, T.} \emph{et~al.}
\newblock \bibinfo{title}{Superradiance in ultracold rydberg gases}.
\newblock \emph{\bibinfo{journal}{Phys. Rev. A}} \textbf{\bibinfo{volume}{75}},
  \bibinfo{pages}{033802} (\bibinfo{year}{2007}).

\bibitem{agarwalana}
\bibinfo{author}{Agarwal, G.~S.}, \bibinfo{author}{Saxena, R.},
  \bibinfo{author}{Narducci, L.~M.}, \bibinfo{author}{Feng, D.~H.} \&
  \bibinfo{author}{Gilmore, R.}
\newblock \bibinfo{title}{Analytical solution for the spectrum of resonance
  fluorescence of a cooperative system of two atoms and the existence of
  additional sidebands}.
\newblock \emph{\bibinfo{journal}{Phys. Rev. A}} \textbf{\bibinfo{volume}{21}},
  \bibinfo{pages}{257--259} (\bibinfo{year}{1980}).

\bibitem{Friedberg1974}
\bibinfo{author}{Friedberg, R.} \& \bibinfo{author}{Hartmann, S.}
\newblock \bibinfo{title}{Superradiant stability in specially shaped small
  samples}.
\newblock \emph{\bibinfo{journal}{Opt. Commun.}}
  \textbf{\bibinfo{volume}{10}}, \bibinfo{pages}{298 -- 301}
  (\bibinfo{year}{1974}).

\bibitem{FriedbergHartman}
\bibinfo{author}{Friedberg, R.} \& \bibinfo{author}{Hartmann, S.~R.}
\newblock \bibinfo{title}{Temporal evolution of superradiance in a small
  sphere}.
\newblock \emph{\bibinfo{journal}{Phys. Rev. A}} \textbf{\bibinfo{volume}{10}},
  \bibinfo{pages}{1728--1739} (\bibinfo{year}{1974}).

\bibitem{Lewenstein}
\bibinfo{author}{Lewenstein, M.} \& \bibinfo{author}{Rzazewski, K.}
\newblock \bibinfo{title}{Collective radiation and the near-zone field}.
\newblock \emph{\bibinfo{journal}{J. Phys. A: Math. Gen.}} \textbf{\bibinfo{volume}{13}},
  \bibinfo{pages}{743--756} (\bibinfo{year}{1980}).

\bibitem{HSteudel1978}
\bibinfo{author}{Steudel, H.} \& \bibinfo{author}{Richter, T.}
\newblock \bibinfo{title}{Radiation properties of a continuously pumped
  two-atom system}.
\newblock \emph{\bibinfo{journal}{Ann. Phys. (Berlin)}}
  \textbf{\bibinfo{volume}{490}}, \bibinfo{pages}{122--136}
  (\bibinfo{year}{1978}).

\bibitem{rzazewski2atom}
\bibinfo{author}{Rza{\.z}ewski, K.} \& \bibinfo{author}{{\.Z}akowicz, W.}
\newblock \bibinfo{title}{Initial value problem for two oscillators interacting
  with electromagnetic field}.
\newblock \emph{\bibinfo{journal}{J. Math. Phys.}}
  \textbf{\bibinfo{volume}{21}}, \bibinfo{pages}{378--388}
  (\bibinfo{year}{1980}).

\bibitem{RuostekoskiPRA1997}
\bibinfo{author}{Ruostekoski, J.} \& \bibinfo{author}{Javanainen, J.}
\newblock \bibinfo{title}{Quantum field theory of cooperative atom response:
  Low light intensity}.
\newblock \emph{\bibinfo{journal}{Phys. Rev. A}} \textbf{\bibinfo{volume}{55}},
  \bibinfo{pages}{513--526} (\bibinfo{year}{1997}).

\bibitem{scullyshift}
\bibinfo{author}{Scully, M.~O.}
\newblock \bibinfo{title}{Collective {L}amb shift in single photon dicke
  superradiance}.
\newblock \emph{\bibinfo{journal}{Phys. Rev. Lett.}}
  \textbf{\bibinfo{volume}{102}}, \bibinfo{pages}{143601}
  (\bibinfo{year}{2009}).

\bibitem{lehmberg}
\bibinfo{author}{Lehmberg, R.~H.}
\newblock \bibinfo{title}{Radiation from an {$N$}-atom system. {I}. {G}eneral
  formalism}.
\newblock \emph{\bibinfo{journal}{Phys. Rev. A}} \textbf{\bibinfo{volume}{2}},
  \bibinfo{pages}{883--888} (\bibinfo{year}{1970}).

\bibitem{dfvjames}
\bibinfo{author}{James, D. F.~V.}
\newblock \bibinfo{title}{Frequency shifts in spontaneous emission from two
  interacting atoms}.
\newblock \emph{\bibinfo{journal}{Phys. Rev. A}} \textbf{\bibinfo{volume}{47}},
  \bibinfo{pages}{1336--1346} (\bibinfo{year}{1993}).


\bibitem{CAdams2012}
\bibinfo{author}{Keaveney, J.} \emph{et~al.}
\newblock \bibinfo{title}{Cooperative lamb shift in an atomic vapor layer of
  nanometer thickness}.
\newblock \emph{\bibinfo{journal}{Phys. Rev. Lett.}}
  \textbf{\bibinfo{volume}{\textbf{108}}}, \bibinfo{pages}{173601}
  (\bibinfo{year}{2012}).

\bibitem{CLS2014}
\bibinfo{author}{Meir, Z.}, \bibinfo{author}{Schwartz, O.},
  \bibinfo{author}{Shahmoon, E.}, \bibinfo{author}{Oron, D.} \&
  \bibinfo{author}{Ozeri, R.}
\newblock \bibinfo{title}{Cooperative lamb shift in a mesoscopic atomic array}.
\newblock \emph{\bibinfo{journal}{Phys. Rev. Lett.}}
  \textbf{\bibinfo{volume}{113}}, \bibinfo{pages}{193002}
  (\bibinfo{year}{2014}).

\bibitem{antoine2015}
\bibinfo{author}{Jennewein, S.}, \bibinfo{author}{Sortais, Y.R.P.},
  \bibinfo{author}{Greffet, J.-J.}\&
  \bibinfo{author}{Browaeys, A.}
\newblock \bibinfo{title}{Propagation of light through small clouds of cold interacting atoms}.
\newblock \emph{\bibinfo{journal}{}}{\bibinfo{volume}{http://arxiv.org/abs/1511.08527}}
  (\bibinfo{year}{2015}).

\bibitem{antoine20152}
\bibinfo{author}{Jennewein, S.}, \bibinfo{author}{Besbes, M.},
  \bibinfo{author}{Schilder, N.J.},\bibinfo{author}{Jenkins, S.D.},
\bibinfo{author}{Sauvan, C.},\bibinfo{author}{Ruostekoski, J.},
\bibinfo{author}{Greffet, J.-J.}, \bibinfo{author}{Sortais, Y.R.P.}\&
  \bibinfo{author}{Browaeys, A.}
\newblock \bibinfo{title}{Observation of the Failure of Lorentz Local field Theory in the Optical Response of Dense and Cold Atomic Systems}.
\newblock \emph{\bibinfo{journal}{}}{\bibinfo{volume}{http://arxiv.org/abs/1510.08041}}
  (\bibinfo{year}{2015}).

\bibitem{antoine2016}
\bibinfo{author}{Jenkins, S.D.}, \bibinfo{author}{Ruostekoski, J.},
\bibinfo{author}{Javanainen, J.},\bibinfo{author}{Bourgain, R.},
\bibinfo{author}{Jennewein, S.},\bibinfo{author}{Sortais, Y.R.P.}\&
  \bibinfo{author}{Browaeys, A.}
\newblock \bibinfo{title}{Optical resonance shifts in the fluorescence imaging of thermal and cold Rubidium atomic gases}.
\newblock \emph{\bibinfo{journal}{}}{\bibinfo{volume}{http://arxiv.org/abs/1602.01037}}
  (\bibinfo{year}{2016}).

\bibitem{CLS2009}
\bibinfo{author}{R{\"o}hlsberger, R.}, \bibinfo{author}{Schlage, K.},
  \bibinfo{author}{Sahoo, B.}, \bibinfo{author}{Couet, S.} \&
  \bibinfo{author}{R{\"u}ffer, R.}
\newblock \bibinfo{title}{Collective lamb shift in single-photon
  superradiance}.
\newblock \emph{\bibinfo{journal}{Science}} \textbf{\bibinfo{volume}{328}},
  \bibinfo{pages}{1248--1251} (\bibinfo{year}{2010}).

\bibitem{Havey2013}
\bibinfo{author}{Balik, S.}, \bibinfo{author}{Win, A.~L.},
  \bibinfo{author}{Havey, M.~D.}, \bibinfo{author}{Sokolov, I.~M.} \&
  \bibinfo{author}{Kupriyanov, D.~V.}
\newblock \bibinfo{title}{Near-resonance light scattering from a high-density
  ultracold atomic ${}^{87}${R}b gas}.
\newblock \emph{\bibinfo{journal}{Phys. Rev. A}} \textbf{\bibinfo{volume}{87}},
  \bibinfo{pages}{053817} (\bibinfo{year}{2013}).

\bibitem{SrFlash2011}
\bibinfo{author}{Chalony, M.}, \bibinfo{author}{Pierrat, R.},
  \bibinfo{author}{Delande, D.} \& \bibinfo{author}{Wilkowski, D.}
\newblock \bibinfo{title}{Coherent flash of light emitted by a cold atomic
  cloud}.
\newblock \emph{\bibinfo{journal}{Phys. Rev. A}} \textbf{\bibinfo{volume}{84}},
  \bibinfo{pages}{011401} (\bibinfo{year}{2011}).

\bibitem{Kaiser2015}
\bibinfo{author}{Guerin, W.}, \bibinfo{author}{Araujo, M. O.},
  \bibinfo{author}{Kaiser, R.}
\newblock \bibinfo{title}{Dicke subradiance in a large cloud of cold atoms}.
\newblock \emph{\bibinfo{journal}{}}{\bibinfo{volume}{http://arxiv.org/abs/1509.00227 \emph{(Phys. Rev. Lett. in press)}}}
(\bibinfo{year}{2015}).

\bibitem{SrFlash2014}
\bibinfo{author}{Kwong, C.~C.} \emph{et~al.}
\newblock \bibinfo{title}{Cooperative emission of a coherent superflash of
  light}.
\newblock \emph{\bibinfo{journal}{Phys. Rev. Lett.}}
  \textbf{\bibinfo{volume}{113}}, \bibinfo{pages}{223601}
  (\bibinfo{year}{2014}).

\bibitem{Kaiser1999}
\bibinfo{author}{Labeyrie, G.} \emph{et~al.}
\newblock \bibinfo{title}{Coherent backscattering of light by cold atoms}.
\newblock \emph{\bibinfo{journal}{Phys. Rev. Lett.}}
  \textbf{\bibinfo{volume}{83}}, \bibinfo{pages}{5266--5269}
  (\bibinfo{year}{1999}).

\bibitem{Havey2003}
\bibinfo{author}{Kulatunga, P.} \emph{et~al.}
\newblock \bibinfo{title}{Measurement of correlated multiple light scattering
  in ultracold atomic ${}^{85}${R}b}.
\newblock \emph{\bibinfo{journal}{Phys. Rev. A}} \textbf{\bibinfo{volume}{68}},
  \bibinfo{pages}{033816} (\bibinfo{year}{2003}).

\bibitem{rmp1998}
\bibinfo{author}{de~Vries, P.}, \bibinfo{author}{van Coevorden, D.~V.} \&
  \bibinfo{author}{Lagendijk, A.}
\newblock \bibinfo{title}{Point scatterers for classical waves}.
\newblock \emph{\bibinfo{journal}{Rev. Mod. Phys.}}
  \textbf{\bibinfo{volume}{70}}, \bibinfo{pages}{447--466}
  (\bibinfo{year}{1998}).

\bibitem{rmp1999}
\bibinfo{author}{van Rossum, M. C.~W.} \& \bibinfo{author}{Nieuwenhuizen,
  T.~M.}
\newblock \bibinfo{title}{Multiple scattering of classical waves: microscopy,
  mesoscopy, and diffusion}.
\newblock \emph{\bibinfo{journal}{Rev. Mod. Phys.}}
  \textbf{\bibinfo{volume}{71}}, \bibinfo{pages}{313--371}
  (\bibinfo{year}{1999}).

\bibitem{Wieman1990}
\bibinfo{author}{Walker, T.}, \bibinfo{author}{Sesko, D.} \&
  \bibinfo{author}{Wieman, C.}
\newblock \bibinfo{title}{Collective behaviour of optically trapped neutral
  atoms}.
\newblock \emph{\bibinfo{journal}{Phys. Rev. Lett.}}
  \textbf{\bibinfo{volume}{64}}, \bibinfo{pages}{408} (\bibinfo{year}{1990}).

\bibitem{Pfau07}
\bibinfo{author}{Heidemann, R.} \emph{et~al.}
\newblock \bibinfo{title}{Evidence for coherent collective {R}ydberg excitation
  in the strong blockade regime}.
\newblock \emph{\bibinfo{journal}{Phys. Rev. Lett.}}
  \textbf{\bibinfo{volume}{99}}, \bibinfo{pages}{163601}
  (\bibinfo{year}{2007}).

\bibitem{Lukin01}
\bibinfo{author}{Lukin, M.~D.} \emph{et~al.}
\newblock \bibinfo{title}{Dipole blockade and quantum information processing in
  mesoscopic atomic ensembles}.
\newblock \emph{\bibinfo{journal}{Phys. Rev. Lett.}}
  \textbf{\bibinfo{volume}{87}}, \bibinfo{pages}{037901}
  (\bibinfo{year}{2001}).

\bibitem{Saffman2010}
\bibinfo{author}{Saffman, M.}, \bibinfo{author}{Walker, T.~G.} \&
  \bibinfo{author}{M\o{}lmer, K.}
\newblock \bibinfo{title}{Quantum information with {R}ydberg atoms}.
\newblock \emph{\bibinfo{journal}{Rev. Mod. Phys.}}
  \textbf{\bibinfo{volume}{82}}, \bibinfo{pages}{2313--2363}
  (\bibinfo{year}{2010}).

\bibitem{Pfau2009}
\bibinfo{author}{Lahaye, T.}, \bibinfo{author}{Menotti, C.}, \bibinfo{author}{Santos, L.}, \bibinfo{author}{Lewenstein, M.} \&
  \bibinfo{author}{Pfau, T.}
\newblock \bibinfo{title}{The physics of dipolar bosonic quantum gases}.
\newblock \emph{\bibinfo{journal}{Rep. Prog. Phys.}}
  \textbf{\bibinfo{volume}{72}}, \bibinfo{pages}{126401}
  (\bibinfo{year}{2009}).

\bibitem{Pfau2012}
\bibinfo{author}{L\"ow, R.} \emph{et~al.}
\newblock \bibinfo{title}{An experimental and theoretical guide to strongly
  interacting {R}ydberg gases}.
\newblock \emph{\bibinfo{journal}{J. Phys. B:At. Mol. Opt. Phys.}}
  \textbf{\bibinfo{volume}{45}}, \bibinfo{pages}{113001}
  (\bibinfo{year}{2012}).

\bibitem{Peyronel2012}
\bibinfo{author}{Peyronel, T.} \emph{et~al.}
\newblock \bibinfo{title}{Quantum nonlinear optics with single photons enabled
  by strongly interacting atoms}.
\newblock \emph{\bibinfo{journal}{Nature}} \textbf{\bibinfo{volume}{488}},
  \bibinfo{pages}{57--60} (\bibinfo{year}{2012}).

\bibitem{Antoine14}
\bibinfo{author}{Ravets, S.} \emph{et~al.}
\newblock \bibinfo{title}{Coherent dipole-dipole coupling between two single
  rydberg atoms at an electrically-tuned f\"orster resonance}.
\newblock \emph{\bibinfo{journal}{Nature Phys.}} \textbf{\bibinfo{volume}{10}},
  \bibinfo{pages}{914--917} (\bibinfo{year}{2014}).

\bibitem{Gunter13}
\bibinfo{author}{G\"unter, G.} \emph{et~al.}
\newblock \bibinfo{title}{Observing the dynamics of dipole-mediated energy
  transport by interaction-enhanced imaging}.
\newblock \emph{\bibinfo{journal}{Science}} \textbf{\bibinfo{volume}{342}},
  \bibinfo{pages}{954--956} (\bibinfo{year}{2013}).


\bibitem{Chang2004}
\bibinfo{author}{Chang, D.~E.}, \bibinfo{author}{Ye, J.} \&
  \bibinfo{author}{Lukin, M.~D.}
\newblock \bibinfo{title}{Controlling dipole-dipole frequency shifts in a
  lattice-based optical atomic clock}.
\newblock \emph{\bibinfo{journal}{Phys. Rev. A}} \textbf{\bibinfo{volume}{69}},
  \bibinfo{pages}{023810} (\bibinfo{year}{2004}).

\bibitem{kus}
\bibinfo{author}{Ku\ifmmode \acute{s}\else \'{s}\fi{}, M. and W\'odkiewicz, K.}
\newblock \bibinfo{title}{Two-atom resonance fluorescence}.
\newblock \emph{\bibinfo{journal}{Phys. Rev. A}}
  \textbf{\bibinfo{volume}{23}}, \bibinfo{pages}{853-857}
  (\bibinfo{year}{1981}).

\bibitem{zakowiczsphere}
\bibinfo{author}{\ifmmode \dot{Z}\else \.{Z}\fi{}akowicz, W.}
\newblock \bibinfo{title}{Superradiant decay of a small spherical system of harmonic oscillators}.
\newblock \emph{\bibinfo{journal}{Phys. Rev. A}}
  \textbf{\bibinfo{volume}{17}}, \bibinfo{pages}{343-352}
  (\bibinfo{year}{1978}).
  
    \bibitem{Sutherland}
\bibinfo{author}{Sutherland, R. T. and Robicheaux, F.}
\newblock \bibinfo{title}{Coherent Forward Broadening in Cold Atom Clouds}.
\newblock \emph{\bibinfo{journal}{Phys. Rev. A}}
  \textbf{\bibinfo{volume}{93}}, \bibinfo{pages}{023407}
  (\bibinfo{year}{2016}).

\bibitem{Nicholson2015}
\bibinfo{author}{Nicholson, T.~L.} \emph{et~al.}
\newblock \bibinfo{title}{Systematic evaluation of an atomic clock at $2 \times
  10^{-18}$ total uncertainty}.
\newblock \emph{\bibinfo{journal}{Nat Commun}} \textbf{\bibinfo{volume}{6}},
  \bibinfo{pages}{6896} (\bibinfo{year}{2015}).
  

\bibitem{Katori2015}
\bibinfo{author}{Ushijima, I.}, \bibinfo{author}{Takamoto, M.},
  \bibinfo{author}{Das, M.}, \bibinfo{author}{Ohkubo, T.} \&
  \bibinfo{author}{Katori, H.}
\newblock \bibinfo{title}{Cryogenic optical lattice clocks}.
\newblock \emph{\bibinfo{journal}{Nat Photon}} \textbf{\bibinfo{volume}{9}},
  \bibinfo{pages}{185--189} (\bibinfo{year}{2015}).

\bibitem{igor}
\bibinfo{author}{Olmos, B.} \emph{et~al.}
\newblock \bibinfo{title}{Long-range interacting many-body systems with
  alkaline-earth-metal atoms}.
\newblock \emph{\bibinfo{journal}{Phys. Rev. Lett.}}
  \textbf{\bibinfo{volume}{110}}, \bibinfo{pages}{143602}
  (\bibinfo{year}{2013}).

\bibitem{Rempe2011}
\bibinfo{author}{Specht, H.~P.} \emph{et~al.}
\newblock \bibinfo{title}{A single-atom quantum memory}.
\newblock \emph{\bibinfo{journal}{Nature}} \textbf{\bibinfo{volume}{473}},
  \bibinfo{pages}{190--193} (\bibinfo{year}{2011}).

\bibitem{juhashift}
\bibinfo{author}{Javanainen, J.}, \bibinfo{author}{Ruostekoski, J.},
  \bibinfo{author}{Li, Y.} \& \bibinfo{author}{Yoo, S.-M.}
\newblock \bibinfo{title}{Shifts of a resonance line in a dense atomic sample}.
\newblock \emph{\bibinfo{journal}{Phys. Rev. Lett.}}
  \textbf{\bibinfo{volume}{112}}, \bibinfo{pages}{113603}
  (\bibinfo{year}{2014}).

\bibitem{thermal}
\bibinfo{author}{Labeyrie, G.}, \bibinfo{author}{Delande, D.},
  \bibinfo{author}{Kaiser, R.} \& \bibinfo{author}{Miniatura, C.}
\newblock \bibinfo{title}{Light transport in cold atoms and thermal
  decoherence}.
\newblock \emph{\bibinfo{journal}{Phys. Rev. Lett.}}
  \textbf{\bibinfo{volume}{97}}, \bibinfo{pages}{013004}
  (\bibinfo{year}{2006}).

\bibitem{Ido2005}
\bibinfo{author}{Ido, T.} \emph{et~al.}
\newblock \bibinfo{title}{Precision spectroscopy and density-dependent
  frequency shifts in ultracold {S}r}.
\newblock \emph{\bibinfo{journal}{Phys. Rev. Lett.}}
  \textbf{\bibinfo{volume}{94}}, \bibinfo{pages}{153001}
  (\bibinfo{year}{2005}).

\bibitem{resonantmultiple}
\bibinfo{author}{Lagendijk, A.} \& \bibinfo{author}{Van~Tiggelen, B.~A.}
\newblock \bibinfo{title}{Resonant multiple scattering of light}.
\newblock \emph{\bibinfo{journal}{Phys. Rep.}} \textbf{\bibinfo{volume}{270}},
  \bibinfo{pages}{143--215} (\bibinfo{year}{1996}).



\bibitem{refindex}
\bibinfo{author}{Morice, O.}, \bibinfo{author}{Castin, Y.} \&
  \bibinfo{author}{Dalibard, J.}
\newblock \bibinfo{title}{Refractive index of a dilute {B}ose gas}.
\newblock \emph{\bibinfo{journal}{Phys. Rev. A}} \textbf{\bibinfo{volume}{51}},
  \bibinfo{pages}{3896--3901} (\bibinfo{year}{1995}).

\bibitem{vector}
\bibinfo{author}{Rouabah, M.-T.} \emph{et~al.}
\newblock \bibinfo{title}{Coherence effects in scattering order expansion of
  light by atomic clouds}.
\newblock \emph{\bibinfo{journal}{J. Opt. Soc. Am. A}}
  \textbf{\bibinfo{volume}{31}}, \bibinfo{pages}{1031--1039}
  (\bibinfo{year}{2014}).

\bibitem{dalibard}
\bibinfo{author}{Chomaz, L.}, \bibinfo{author}{Corman, L.},
  \bibinfo{author}{Yefsah, T.}, \bibinfo{author}{Desbuquois, R.} \&
  \bibinfo{author}{Dalibard, J.}
\newblock \bibinfo{title}{Absorption imaging of a quasi-two-dimensional gas: a
  multiple scattering analysis}.
\newblock \emph{\bibinfo{journal}{New J. Phys.}} \textbf{\bibinfo{volume}{14}},
  \bibinfo{pages}{055001} (\bibinfo{year}{2012}).

\bibitem{bienaime2011}
\bibinfo{author}{Bienaim\'e, T.}, \bibinfo{author}{Petruzzo, M.},
  \bibinfo{author}{Bigerni, D.}, \bibinfo{author}{Piovella, N.} \&
  \bibinfo{author}{Kaiser, R.}
\newblock \bibinfo{title}{Atom and photon measurement in cooperative scattering
  by cold atoms}.
\newblock \emph{\bibinfo{journal}{J. Mod. Opt.}} \textbf{\bibinfo{volume}{58}},
  \bibinfo{pages}{1942--1950} (\bibinfo{year}{2011}).

\bibitem{bienaime2013}
\bibinfo{author}{Bienaim\'e, T.}, \bibinfo{author}{Bachelard, R.},
  \bibinfo{author}{Piovella, N.} \& \bibinfo{author}{Kaiser, R.}
\newblock \bibinfo{title}{Cooperativity in light scattering by cold atoms}.
\newblock \emph{\bibinfo{journal}{Fortschr. Physik}}
  \textbf{\bibinfo{volume}{61}}, \bibinfo{pages}{377--392}
  (\bibinfo{year}{2013}).

\bibitem{opticaltheo}
\bibinfo{author}{Bienaim\'e, T.} \emph{et~al.}
\newblock \bibinfo{title}{Interplay between radiation pressure force and
  scattered light intensity in the cooperative scattering by cold atoms}.
\newblock \emph{\bibinfo{journal}{J. Mod. Opt.}} \textbf{\bibinfo{volume}{61}},
  \bibinfo{pages}{18--24} (\bibinfo{year}{2014}).

\bibitem{cord2002}
\bibinfo{author}{M\"uller, C.~A.} \& \bibinfo{author}{Miniatura, C.}
\newblock \bibinfo{title}{Multiple scattering of light by atoms with internal
  degeneracy}.
\newblock \emph{\bibinfo{journal}{J. Phys. A: Math. Gen.}}
  \textbf{\bibinfo{volume}{35}}, \bibinfo{pages}{10163} (\bibinfo{year}{2002}).



\bibitem{Eberly}
\bibinfo{author}{Allen, L.} \& \bibinfo{author}{Eberly, J.~H.}
\newblock \emph{\bibinfo{title}{Optical Resonance and Two-Level Atoms}}.
\newblock Dover Books on Physics (\bibinfo{publisher}{Dover Publications},
  \bibinfo{year}{1987}).

\bibitem{Pellegrino2014}
\bibinfo{author}{Pellegrino, J.} \emph{et~al.}
\newblock \bibinfo{title}{Observation of suppression of light scattering
  induced by dipole-dipole interactions in a cold-atom ensemble}.
\newblock \emph{\bibinfo{journal}{Phys. Rev. Lett.}}
  \textbf{\bibinfo{volume}{113}}, \bibinfo{pages}{133602}
  (\bibinfo{year}{2014}).

\bibitem{Yelin99}
\bibinfo{author}{Fleischhauer, M.} \& \bibinfo{author}{Yelin, S.~F.}
\newblock \bibinfo{title}{Radiative atom-atom interactions in optically dense
  media: quantum corrections to the lorentz-lorenz formula}.
\newblock \emph{\bibinfo{journal}{Phys. Rev. A}} \textbf{\bibinfo{volume}{59}},
  \bibinfo{pages}{2427} (\bibinfo{year}{1999}).

\bibitem{Yelin2012}
\bibinfo{author}{Lin, G.-D.} \& \bibinfo{author}{Yelin, S.~F.}
\newblock \bibinfo{title}{Chapter 6 - superradiance: An integrated approach to
  cooperative effects in various systems}.
\newblock In \bibinfo{editor}{Paul~Berman, E.~A.} \& \bibinfo{editor}{Lin, C.}
  (eds.) \emph{\bibinfo{booktitle}{Advances in Atomic, Molecular, and Optical
  Physics}}, {V}ol.~\bibinfo{volume}{61} of \emph{\bibinfo{series}{Advances In
  Atomic, Molecular, and Optical Physics}}, \bibinfo{pages}{295 -- 329}
  (\bibinfo{publisher}{Academic Press}, \bibinfo{year}{2012}).

\bibitem{freqshift}
\bibinfo{author}{Friedberg, R.}, \bibinfo{author}{Hartmann, S.} \&
  \bibinfo{author}{Manassah, J.}
\newblock \bibinfo{title}{Frequency shifts in emission and absorption by
  resonant systems ot two-level atoms}.
\newblock \emph{\bibinfo{journal}{Phys. Rep.}} \textbf{\bibinfo{volume}{7}},
  \bibinfo{pages}{101 -- 179} (\bibinfo{year}{1973}).

\bibitem{Julienne1996}
\bibinfo{author}{Julienne, P.~S.}
\newblock \bibinfo{title}{Cold Binary Atomic Collisions in a Light Field}.
\newblock \emph{\bibinfo{journal}{J. Res. Natl. Inst. Stand. Technol.}}
  \textbf{\bibinfo{volume}{101}}, \bibinfo{pages}{487}
  (\bibinfo{year}{1996}).


\end{thebibliography}

\begin{thebibliography}{1}
\expandafter\ifx\csname url\endcsname\relax
  \def\url#1{\texttt{#1}}\fi
\expandafter\ifx\csname urlprefix\endcsname\relax\def\urlprefix{URL }\fi
\providecommand{\bibinfo}[2]{#2}
\providecommand{\eprint}[2][]{\url{#2}}

\bibitem{Ido2005}
\bibinfo{author}{Ido, T.} \emph{et~al.}
\newblock \bibinfo{title}{Precision spectroscopy and density-dependent
  frequency shifts in ultracold {S}r}.
\newblock \emph{\bibinfo{journal}{Phys. Rev. Lett.}}
  \textbf{\bibinfo{volume}{94}}, \bibinfo{pages}{153001}
  (\bibinfo{year}{2005}).

\bibitem{dalibard}
\bibinfo{author}{Chomaz, L.}, \bibinfo{author}{Corman, L.},
  \bibinfo{author}{Yefsah, T.}, \bibinfo{author}{Desbuquois, R.} \&
  \bibinfo{author}{Dalibard, J.}
\newblock \bibinfo{title}{Absorption imaging of a quasi-two-dimensional gas: a
  multiple scattering analysis}.
\newblock \emph{\bibinfo{journal}{New J. Phys.}} \textbf{\bibinfo{volume}{14}},
  \bibinfo{pages}{055001} (\bibinfo{year}{2012}).

\bibitem{subradiance}
\bibinfo{author}{Bienaim\'e, T.}, \bibinfo{author}{Piovella, N.} \&
  \bibinfo{author}{Kaiser, R.}
\newblock \bibinfo{title}{Controlled dicke subradiance from a large cloud of
  two-level systems}.
\newblock \emph{\bibinfo{journal}{Phys. Rev. Lett.}}
  \textbf{\bibinfo{volume}{108}}, \bibinfo{pages}{123602}
  (\bibinfo{year}{2012}).

\bibitem{juhashift}
\bibinfo{author}{Javanainen, J.}, \bibinfo{author}{Ruostekoski, J.},
  \bibinfo{author}{Li, Y.} \& \bibinfo{author}{Yoo, S.-M.}
\newblock \bibinfo{title}{Shifts of a resonance line in a dense atomic sample}.
\newblock \emph{\bibinfo{journal}{Phys. Rev. Lett.}}
  \textbf{\bibinfo{volume}{112}}, \bibinfo{pages}{113603}
  (\bibinfo{year}{2014}).

\bibitem{Burnett1992}
\bibinfo{author}{Trippenbach, M.}, \bibinfo{author}{Gao, B.},
  \bibinfo{author}{Cooper, J.} \& \bibinfo{author}{Burnett, K.}
\newblock \bibinfo{title}{Slow collisions between identical atoms in a laser
  field: The spectrum of redistributed light}.
\newblock \emph{\bibinfo{journal}{Phys. Rev. A}} \textbf{\bibinfo{volume}{45}},
  \bibinfo{pages}{6555--6569} (\bibinfo{year}{1992}).

\end{thebibliography}

\begin{addendum}
\item[Acknowledgments] We are grateful to Paul Julienne, Chris Greene, John Cooper, and Murray Holland for their important insights and stimulating discussions.  This research is supported by NIST, NSF Physics Frontier Center at JILA, AFOSR, AFOSR-MURI, ARO, DARPA QuASAR, NSF Center for Ultracold Atoms at Harvard-MIT, ITAMP, and ANR-14-CE26-0032.

\item[Author contributions] S.L.B., M.B., X.Z. T.B. T.L.N., and J.Y. contributed to the executions of the experiments.  B.Z., J.S., R.K., S.F.Y., M.D.L. and A.M.R. contributed to the development of the theory model.  All authors discussed the results, contributed to the data analysis, and worked together on the manuscript.

\item[Author information] Reprints and permissions information is available at www.nature.com/reprints. The authors have no competing financial interest. Correspondence and requests for materials should be addressed to A.M.R. (arey@jilau1.colorado.edu) or J.Y. (Ye@jila.colorado.edu).

 \item[Competing financial interests] The authors declare that they have no competing financial interests.
\end{addendum}
\clearpage
\clearpage
\clearpage

\clearpage
\clearpage
\newpage
\begin{center}
\textbf{\Large Supplementary Information}\\*
\end{center}

\setcounter{equation}{0}
\noindent{\bf Supplementary Figures}
\vspace{-0.6cm}
\\
\begin{suppfigure}[H]
\centering
\includegraphics[width=0.5\textwidth]{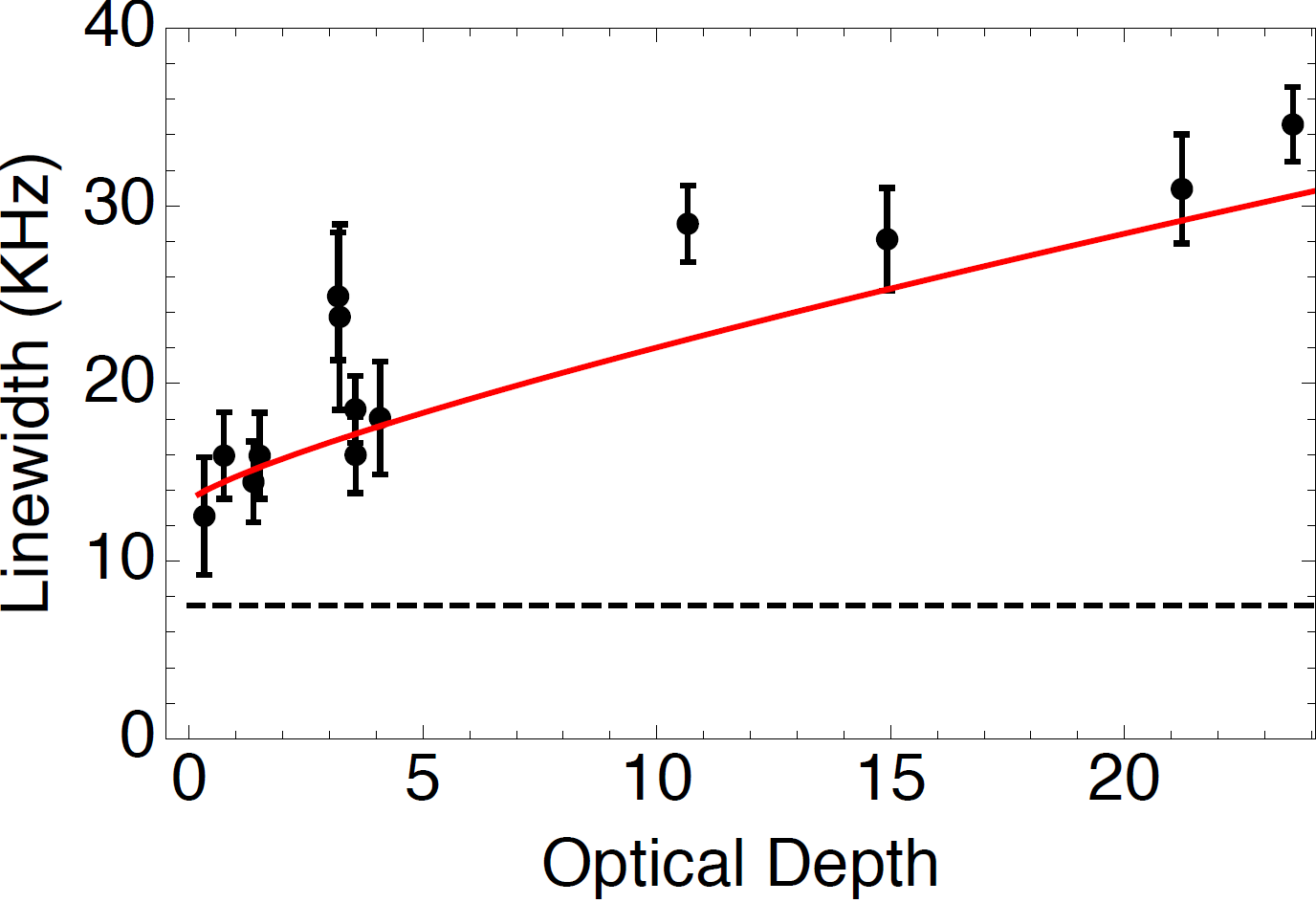}
  \caption{ \label{subfig:linewidth}  {\bf Comparison between theoretical calculations and the linewidth data in Ref. [\citenum{Ido2005}]}. The linewidth data were taken under the same condition as the red data in Fig. 4, where motional effects are significant (see Ref. [\citenum{Ido2005}] for experimental details). Still the theoretical model presented here can capture the linewidth broadening. Here $\eta=1.5$ is used in the numerical simulations. All experimental errorbars are for statistical uncertainties.}
\end{suppfigure}

\noindent{\bf Supplementary Note 1: On resonance optical depth}\\
\vspace{-0.6cm}
\\
For the $J=0\rightarrow J=1$ transition, the  atom-photon scattering cross section is $\mathit{\Sigma}(\mathit{\Delta})=\frac{6\pi}{k^2}\frac{1}{1+4(\mathit{\Delta}/\mathit{\Gamma})^2}$, with $k$ the wavevector of the photon and $\mathit{\Delta}$, $\mathit{\Gamma}$ are the detuning of the driving laser and the natural linewidth, respectively~\cite{dalibard}.  In the experiment, the atomic cloud has approximately a Gaussian distribution $n(x,y,z)=n_0e^{-\frac{x^2}{2R_x^2}-\frac{y^2}{2R_y^2}-\frac{z^2}{2R_z^2}}$, where $n_0$ satisfies $\int dxdydz n(x,y,z)=N$, and $N$ is the total number of atoms. Along the line of observation, {\em e.g.} $\hat{x}$, the on resonance optical depth is related to the resonant scattering cross section $\mathit{\Sigma}_0=\frac{6\pi}{k^2}$, and the column density averaged over the profile perpendicular to this direction~\cite{dalibard,subradiance},
\begin{eqnarray}
OD&=&[\int dy dz n(y,z)]^{-1}\int dy dz  n(y,z) OD(y,z)\\
&=&[\int dy dz n(y,z)]^{-1}\int dy dz n(y,z)\int dx n(x,y,z)\mathit{\Sigma}_0\\
&=&[\int dy dz n(y,z)]^{-1}\int dy dz n(y,z)e^{-\frac{y^2}{2R_y^2}-\frac{z^2}{2R_z^2}}\int dx n_0e^{-\frac{x^2}{2R_x^2}}\mathit{\Sigma}_0\\
&=&\frac{3N}{2k^2R_yR_z}\\
&=&\frac{3N}{2k^2R_{\perp}^2},
\end{eqnarray}
which is the $OD$ defined in the maintext.

\noindent{\bf Supplementary Note 2:  Numerical simulation of the coherent dipole model}\\
\vspace{-0.6cm}
\\
To simulate the experiment, we use $N_{\rm sim}\sim3000$ to $10000$ atoms and assume they are  randomly distributed according to a  density distribution $n(x,y,z)\propto e^{-\frac{x^2}{2\sigma_x^2}-\frac{y^2}{2\sigma_y^2}-\frac{z^2}{2\sigma_z^2}}$, where $\sigma_{x,y,z}$ denote the widths of the atomic cloud in the simulation, and the aspect  $\sigma_x:\sigma_y:\sigma_z$ is kept the same as the one measured in experiment. We average over a sufficient amount of configurations until  convergence is achieved.   To reproduce the behavior of the  $N_{\rm exp}\sim10^7$ Sr atoms interrogated in the  experiment, with  the  smaller number of atoms used in the numerical simulations we need to rescale the  widths $\sigma_\alpha$. For linewidth and fluorescence intensity, which are dominantly $OD$ effects, the appropriate rescaling would be to match the $OD$ used in the experiment and require  $\sigma_{\perp}^{OD}=(N_{\rm sim}/N_{\rm exp})^{1/2} R_\perp$. However, under that procedure the density used in the theory does not match the experimental densities, instead it is much larger due to the factor  $N_{\rm exp}/N_{\rm sim}\sim 10^4$, and this introduces non-negligible modification on the linewidth. We find that  this issue can be compensated  by a  constant rescaling  of the width  $\sigma^{\rm sim}_\perp=\eta \sigma_{ OD}$.  In the numerical simulation, we keep the  parameter $\eta$  constant to model all the experimental measurements taken under the same conditions. In Fig. 2(a) and (b), we set $\eta=5$ for both the blue and red probe simulations under  different detection angles $\theta$ and atom numbers. The experimental measurements shown in Fig. 2(c) and Fig. 3(a) (b) were taken under different geometries, and we use  $\eta=2.35$ for those simulations.

We account for  motional effects in the  red transition by introducing  random detunings $\delta \nu$ for each atom and   these are sampled according to a gaussian thermal distribution 
\begin{equation}
P(\delta\nu)=\frac{1}{\sqrt{2\pi}\mathit{\Delta}_{\rm D}}{\rm exp}\left(-\frac{\delta \nu^2}{2 \mathit{\Delta}_{\rm D}^2}\right).
\end{equation}
 Here $\mathit{\Delta}_{\rm D}$ is the Doppler width at the experimental temperature. Specifically, for non-interacting two-level atoms, the atomic coherence is modified as
\begin{eqnarray}
b_j&=&\frac{\Omega e^{i{\bf k}\cdot {\bf r}_j}}{(\mathit{\Delta}-\delta\nu_j)+i\mathit{\Gamma}/2}.
\end{eqnarray}
For the incoherent scattering, this leads to
\begin{eqnarray}
I_{\rm incoh}&=&\frac{1}{\sqrt{2\pi}\mathit{\Delta}_{\rm D}}\int d\delta\nu_j |b_j|^2e^{-\delta\nu_j^2/2\mathit{\Delta}_{\rm D}^2},
\end{eqnarray}
 while for the coherent scattering in the forward direction, one has to take into account pairwise atomic contributions, such that
\begin{eqnarray}
I_{\rm coh}&=&\frac{1}{2\pi\mathit{\Delta}_{\rm D}^2}\int d\delta\nu_{j}d\delta\nu_{j'} b_{j}b^*_{j'}e^{-\delta\nu_{j}^2/2\mathit{\Delta}_{\rm D}^2}e^{-\delta\nu_{j'}^2/2\mathit{\Delta}_{\rm D}^2},
\end{eqnarray}
thus the on-resonance enhancement factor is
\begin{eqnarray}
\frac{I_{\rm coh}}{I_{\rm incoh}}&=&\frac{\sqrt{\frac{\pi}{2}}e^{\frac{1}{8\mathit{\Delta}_{\rm D}^2/\mathit{\Gamma}^2}}{\rm Erfc}(\frac{1}{2\sqrt{2}\mathit{\Delta}_{\rm D}/\mathit{\Gamma}})}{2\mathit{\Delta}_{\rm D}/\mathit{\Gamma}},
\end{eqnarray}
where ${\rm Erfc}$ is the complementary error function. This shows a suppression of the forward interference that depends on $\mathit{\Delta}_{\rm D}/\mathit{\Gamma}$.

A similar procedure was used in Refs.~[\citenum{subradiance,juhashift}] to deal with motional effects. This simple treatment of the atomic motion  accounts only for the  Doppler shifts experienced by the atoms, but neglects light induced mechanical effects on atoms and random phase evolution in the dipole-dipole coupling due to atomic motion \cite{subradiance}. We expect it to be  valid  if the atoms are weakly driven, $\Omega\ll \mathit{\Gamma}$, and  for low velocities $k v\leq \mathit{\Gamma}$ or short probing times $t\lesssim 1/kv$. Those conditions are more or less satisfied    in   the red measurements presented in Figs. 2 and 3.
  The  random dephasing added in  the coherent dipole model  gives rise to  a Voigt profile lineshape with a constant  Gaussian width that is consistent with $\mathit{\Delta}_{\rm D}$, and with a Lorentzian width that increases with $OD$. In Fig. 3(c), an  $\eta=2.35$ is used in the simulations for all the different  $OD$ conditions.

For treating the density shift, which is  dominantly a density effect, it is more appropriate to rescale the  cloud size to match the experimental density, $\bar{\sigma}^{\rm{density}}=(N_{\rm sim}/N_{\rm exp})^{1/3} \bar{R}$. Here $\bar{\sigma}$ and  $\bar{R}$  are the corresponding geometric means. This is the procedure  we use to produce the theory data presented in Fig. 4.  In this case, nevertheless, the frozen dipole approximation is not able to capture the large density shift observed in the red probe experiment. The model is only able to reproduce the blue probe density shift.  The red frequency shift measurements were performed under different conditions than those used for Figs. 2 and 3 (see Ref.~[\citenum{Ido2005}]). The failure of the frozen dipole model to reproduce the density shift in situations  when  atoms are allowed to  be a significant amount of  time in the excited state  and closely approach to each other in a collision event,  emphasizes the need to fully model the interplay between  short and long ranged interactions, and atom motion in a dense sample  \cite{Burnett1992}. While such effects are crucial for the frequency shift in atomic emission \cite{juhashift}, in our calculations, the collective linewidth broadening turns out to be less affected. In Supplementary Fig. 1, the numerical results obtained with the same procedure as described above are compared with the red transition linewidth data measured together with the density shift (see Fig. 4) in Ref. [\citenum{Ido2005}]. The fair agreement between theory and experiment suggests that, even when motion is important, the frozen dipole model is capable of capturing some of the relevant features of collective atomic emission in a dense medium.

\noindent{\bf Supplementary References}

\end{document}